\documentclass{aa}    %% Astronomy & Astrophysics style class aa.cls v8.2

\usepackage{url,twoopt}
\usepackage{graphicx}
\usepackage{graphics}
\usepackage[varg]{txfonts}           %% old-fashioned A&A font choice
\usepackage{hyperref}
\usepackage{array}
\usepackage{multirow}
\usepackage{color}

\begin{document}

\title{Super-Eddington accretion in the Q2237+0305 quasar?}

\author{L.A.Berdina\inst{\ref{inst1},\ref{inst2}}
\and V.S.Tsvetkova\inst{\ref{inst1},\ref{inst2}}
\and V.M.Shulga\inst{\ref{inst1},\ref{inst3},\ref{inst4}}}

\institute{Institute of Radio Astronomy of the National Academy of Sciences of Ukraine, 4 Mystetstv, 61002 Kharkov, Ukraine
\\ \email{lberdina@rian.kharkov.ua; tsvet999@gmail.com; shulga@rian.kharkov.ua}\label{inst1}
\and Institute of Astronomy of Kharkov National University, Sumskaya 35, 61022 Kharkov, Ukraine\label{inst2}
 \and V.N.Karazin Kharkov National University, Svobody sq. 4, 61070 Kharkov, Ukraine\label{inst3}
  \and College of Physics, International Center of Future Science, Jilin University, 2699 Qianjin St., 130012 Changchun, China \label{inst4} }

\date{Received .. /
Accepted ..}

\abstract{ The interband time lags between the flux variations of the Q2237+0305 quasar have been
determined from light curves in the Johnson-Cousins $V$, $R,$ and $I$ spectral bands. The values
of the time lags for filter pairs $R-V$, $I-R,$ and $I-V$ are significantly higher than those predicted
by the standard accretion disk model by Shakura and Sunyaev. To explain the discrepancy,
the idea of a supercritical accretion regime in quasars considered in 1973 by Shakura
and Sunyaev is applied. This regime has been shown by them to cause an extended scattering envelope around the accretion disk. The envelope
efficiently scatters and re-emits the radiation from the accretion disk
and thus increases the apparent disk size. We made use of analytical expressions for the
envelope radius and temperature derived by Shakura and Sunyaev in their analysis of super-Eddington accretion and show that our results
are consistent with the existence of such an envelope. The corresponding parameters of the accretion regime were calculated. They provide the radii of the envelope in the $V$, $R,$ and $I$ spectral bands consistent with the inter-band time lags
determined in our work.}


\titlerunning{Q2237+0305 accretion disk structure}
\authorrunning{Berdina L.A. et al.}
\maketitle

\section{Introduction}

Flux variability inherent in active galactic nuclei (AGNs) and quasars is an important source of information about their physical properties and therefore has been closely investigated by astronomers for decades (see, e.g., Cristiani et al. 1997; Giveon et al.
1999; Vanden Berk et al. 2004; Wilhite et al. 2005;
Magdis \& Papadakis 2006; Gopal-Krishna et al. 2013;  Kumar et al. 2015; Grier et al. 2019; Yi et al. 2019;  Kokubo \& Minezaki 2020; Luo et al. 2020). The variability is observed in all spectral regions, from the X-rays to radio wavelengths, and at timescales from a few days to several years (Giveon et al. 1999; Webb \& Malkan 2000; Wilhite et al. 2005; Schmidt et al. 2012). The characteristic amplitudes grow toward longer time lags and shorter wavelengths.
The flux variations in AGNs and quasars in different spectral bands are often observed with certain time lags between them. This may suggest that these time lags measure the light travel times between the quasar regions that radiate in different spectral bands. It may therefore be a useful instrument for studying spatial structure and physical parameters of AGNs and quasars. This instrument, called ”reverberation mapping” (RM), was initially proposed for measuring the distance between the central region of a quasar that is responsible for the hard continuum radiation and a broad emission line region (Blandford \& McKee 1982). Initially, RM implied spectroscopic observations. The so-called photometric RM has been widely used recently (e.g. Bachev 2009; Edri et al 2012; Jiang et al. 2016, Mudd et al. 2018; Kokubo 2018 ). The photometric RM implied photometry in two or more spectral bands. Some of them contained emission lines and others did not.

In 1997, Wanders et al. (1997) reported reverberation time delays measured
in their spectroscopic RM campaign with the International Ultraviolet Explorer (IUE). The authors noted that the most remarkable result is the detection of apparent time shifts between the brightness variations in different regions of the UV continuum. Collier et al. (1999) later realized that the optical continuum variations lag the UV variations in NGC 7469. Further photometric RM projects provided much
evidence of lags between flux variations observed in continua of
different spectral regions (e.g., Collier 2001; Sergeev et al. 2005; Cackett et al. 2007;
Bachev 2009;  Fausnaugh et al. 2016; Fausnaugh et al. 2018; Mudd et al. 2018). Currently,
several major international projects are dedicated to RM of quasars and AGNs, such as the Sloan Digital Sky Survey Reverberation Mapping (SDSS-RM), the Space
Telescope and Optical Reverberation Mapping (STORM), the Lick AGN Monitoring Project (LAMP), the Dark Energy Survey (DES), and others (Grier et al. 2017; Fausnaugh et al. 2018; Mudd et al. 2018; Homayouni et al. 2019; Kinemuchi et al. 2020; Yu et al. 2020).

The UV/optical radiation from quasars is generally believed to
come from a geometrically thin optically thick accretion disk with a supermassive black
hole in its center. According to this model (Shakura \& Sunyaev 1973), the accretion
disk temperature varies along the disk radius as $T\propto r^{-3/4}$, therefore
the shorter-wavelength radiation must originate closer to the accretion disk center. According to the continuum thermal reprocessing scenario proposed by Krolik et al. (1991), its variations must therefore precede those observed at longer wavelengths. The expected
trend of time lags with wavelength must therefore follow the relationship $\tau\propto \lambda^{4/3}$, provided the accretion disk radiates as a blackbody.

The paper contains photometric RM results and their analysis applied to the $V$, $R,$ and $I$  light curves of Q2237+0305 (nicknamed “the Einstein Cross”, discovered by Huchra et al. in 1985), which is one of the most famous gravitational lens systems with a radio-quiet quasar at $z_q=1.695$  that is quadruply lensed by a $z_g=0.039$ Sab galaxy. Four individual macroimages labeled A through D are arranged in a cross-like pattern around the galaxy nucleus within a circle of approximately 1.8 arcsec in diameter.
Until recently, there were two determinations of the interband time delays for a quasar of the Q2237+0305 lens system (Koptelova et al. 2006; Koptelova et al. 2010). In both works a cross-correlation analysis was applied to the light curves in filter $V$ obtained by the OGLE collaboration (Udalski et al. 2006) and the $R$ light curves built from observations at the 1.5-meter telescope of the Maidanak observatory. The time delays between the variations in these two spectral bands were reported in Koptelova et al. (2006) to be 9 days for image A and 16.2 days for image C. Koptelova et al. (2010) reported time lags of 5.1-5.6 days and 5.1-5.2 days for images A and C, respectively. In addition to different sampling rates and time coverage, the two datasets used in the two works were processed with different photometry algorithms, therefore we decided to repeat the work with more homogeneous data. To do this, we turned to the monitoring data of the Q2237 + 0305 system in filters $V$, $R,$ and $I$ in 2001-2008, presented by Dudinov et al. (2011). Preliminary processing of the 2004 - 2005 light curves in filters $V$ and $R$ has definitely shown that the variations in filter $R$ lag those in $V$ by about 5 days (Berdina et al. 2018, 2019).  We show the results of processing the data of two seasons in all the three filters, $V$, $R,$ and $I$, with a more careful analysis of errors and a subsequent discussion of possible mechanisms leading to the particular values of the interband time delays and their behavior in wavelengths.

In the next section, we describe the initial data we used and discuss their suitability for determining the interband time delays in Q2237+0305.

\section {Initial data and time lags}

Of all the existing monitoring data for Q2237+0305, the detailed light curves in filter $V$ obtained in the framework of the OGLE program (Udalski et al. 2006) are the most famous. Their high photometric accuracy, rather good sampling and time coverage made the OGLE light curves a basis for various microlensing studies, which provided many estimates of the Q2237+0305 lens parameters.

\subsection {Initial data}

\begin{figure}
\resizebox{8.8 cm}{!}{\includegraphics{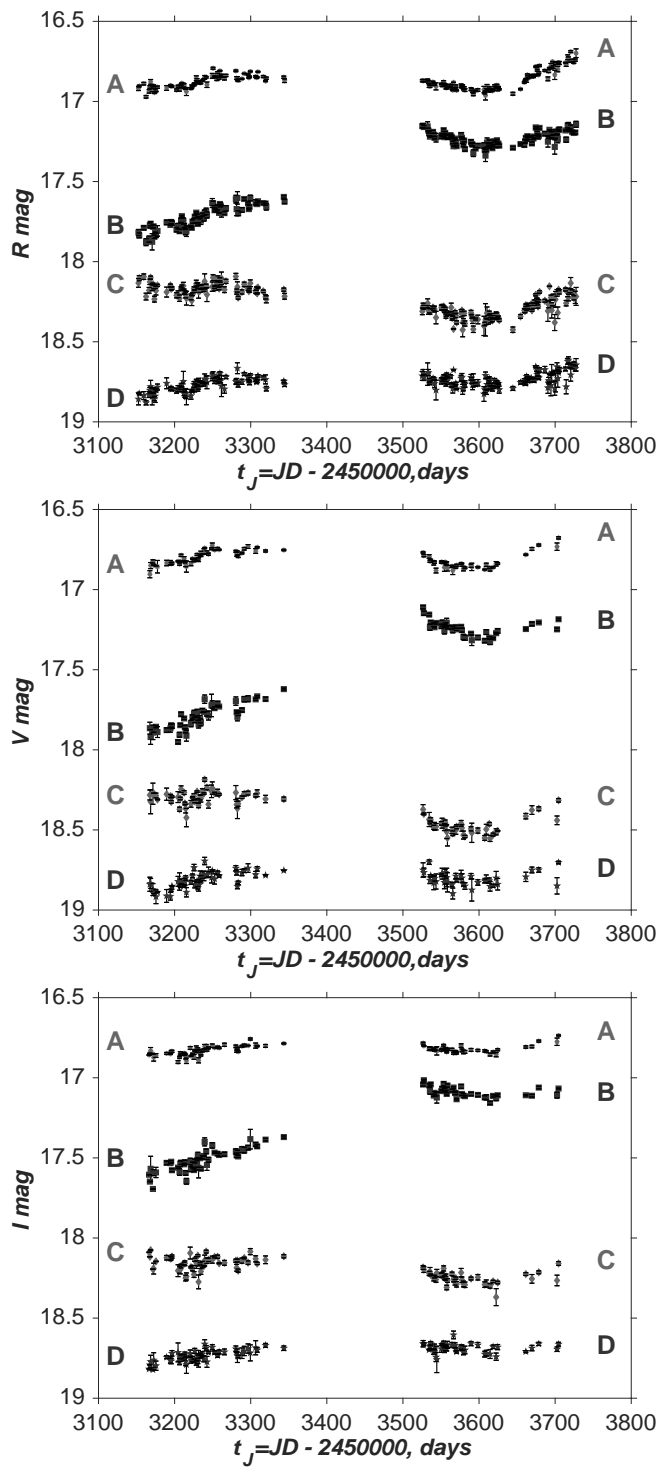}}
\caption{ Light curves of Q2237+0305 in filters $V$, $R,$ and $I$ from observations by Dudinov et al. 2011. The data for the seasons in 2004 and 2005 are presented.}
\label{light curves}
\end{figure}

A similar dataset exists that is much less well known because it was published in a
poorly accessible journal (Dudinov et al. 2011). While the OGLE data are more
accurate and cover a longer observing seasons, the data by Dudinov et al. (2011) have
the advantage that they have been obtained through three filters, -- $V$, $R,$ and $I$
of the Johnson-Cousins photometric system. The effective wavelengths are $\lambda_{eff}
(V)=545.7$ nm, $\lambda_{eff}(R)=634.9$ nm, and $\lambda_{eff}(I)=879.7$ nm, which correspond to $\lambda_{rest}(V)=203.2$ nm, $\lambda_{rest}(R)=
235.6$ nm, and $\lambda_{rest}(I)=326.4$ nm in the Q2237+0305 quasar source plane ($z_q=1.695$).
Photometry spanning eight years, from 2001 to 2008, is available at
the site of the IA of the V.N. Karazin National University (http://www.astron.kharkov.ua/databases/index.html). In Fig. \ref{light curves} we reproduce the light curves of Q2237+0305 for the 2004 and 2005 seasons built from the data by Dudinov et al. (2011). General information about the $V$, $R,$ and $I$ light curves for the 2004 and 2005 seasons is given in Table \ref{data observations}.

%***************************Figure 1******************************************************************
%***********************************************************************************************************

The light curves by Dudinov et al. (2011) and those in filter $R$ used by Koptelova
et al. (2006; 2010) were obtained from the same monitoring data taken with the
1.5-meter telescope of the Maidanak Observatory, but they were processed with different
photometry algorithms, as described in Vakulik et al. (2004), Dudinov et al. (2011), and Koptelova et al. (2005).

%*****************Table 1 ******************************************************************************
\begin{table*} [!htp]
\caption {General information about the data of observations of Q2237+0305 used to determine the interband time delays in this work: duration of seasons, number of data points, photometry error, and seeing (full width at half maximum) averaged over each season.}
\label{data observations}
\centering
\setlength \extrarowheight{2pt}
\begin{tabular}{lcccc}
\hline
  \multirow{2}{*}{Filter} & Season duration &  Number of data  &   \multirow{2}{*}{Photometry error}  &   \multirow{2}{*}{FWHM} \\
 &(days) &  points &   &   \\ [2pt]
\hline
$R$, season 2004 &  192   & 75 & 0.0142 & 1.0475 \\
$R$, season 2005   & 202 & 98 & 0.0141 & -- \\
\hline
$V$, season 2004 & 176  & 48 & 0.0169& 1.0658 \\
$V$, season 2005   & 178 & 38 & 0.0170 & 1.0945\\
\hline
$I$, season 2004 & 177  & 52 & 0.0149& 0.9842 \\
$I$, season 2005   & 178 & 39 & 0.0128& 0.9938 \\
\hline
\end{tabular}
\end{table*}
%**************************************************************************************************

Measurements of the interband time delays in gravitationally lensed quasars encounter almost the same difficulties as those that are inherent in measuring the single-band time delays between the quasar macroimages. These are listed and analyzed in detail by Tewes et al. (2013), for instance. In addition to the known difficulties, which are common for all time-delay lenses (finite photometric accuracy, poor and irregular sampling, gaps between the seasons, variable microlensing), the microlensing events in Q2237+0305 are very frequent; the amplitudes vary from several hundredths of a magnitude to about 1 mag (see the light curves by Udalski et al. (2006) and Dudinov et al. (2011)). Dudinov et al. (2011) estimated the contributions from the source intrinsic variability and microlensing variability into the total flux variations of the Q2237+0305 macroimages during 2000-2008. The two constituents of variability were found to contribute almost equally to the root mean square (RMS) variation of the Q2237+0305 light curves  in 2001-2008.

\begin{figure*} [!htp]
\resizebox{18.3 cm}{!}{\includegraphics{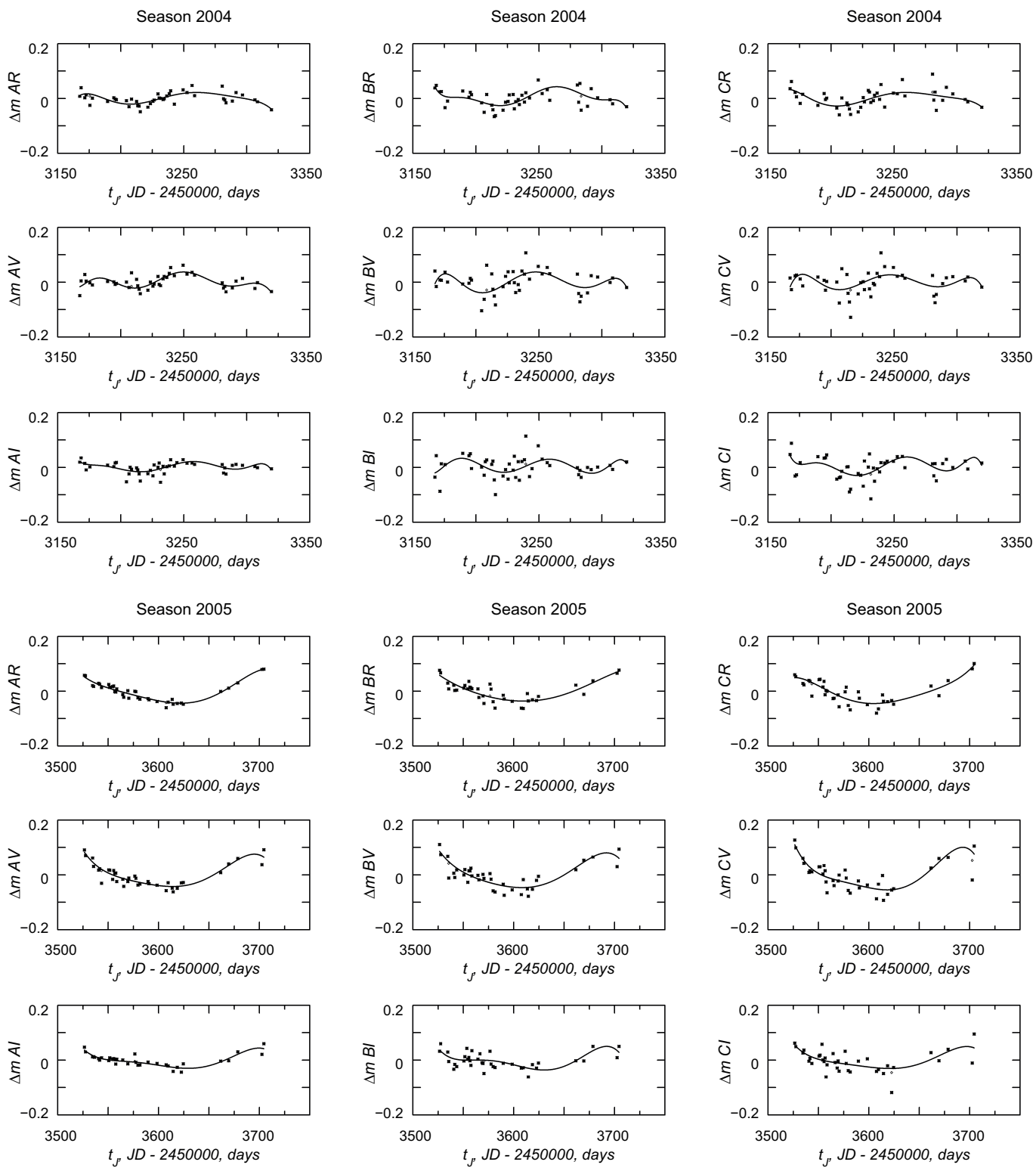}}
 \caption{ Polynomial approximations for the light curves in Fig. \ref{light curves}: the curves are reduced to the same magnitude level, and the first-order terms are eliminated ($ \Delta m $ is the brightness variations in magnitude relative to the average level for a given season). }
\label{polynomial approximations}
\end{figure*}

As microlensing events occur in each macroimage independently, they may cause dramatic mutual distortions of their light curves. An adequate elimination of microlensing effects is therefore vital to obtain the unbiased estimates of the time delays between macroimages. A pair of light curves in two different filters of the same macroimage can be anticipated to be less mutually distorted by microlensing: observations show only a smooth growth of microlensing amplitudes toward the shorter wavelengths, but the effects of microlensing still have to be eliminated. To measure the interband time delays, we therefore applied the method developed by us earlier for determining the differential time delays in gravitationally lensed quasars in the presence of microlensing. In short, the method uses some useful properties of representing the data of observations by the orthogonal polynomials, which provide certain simplicity and convenience in calculations. In particular, it allows any term of the polynomial that approximates a particular light curve to be eliminated or added again without requiring to recalculate the remaining expansion coefficients. This property is useful, in particular, to determine the time delays in the presence of microlensing. A detailed description of the method can be found in Tsvetkova et al. (2016), where its application is
demonstrated in determining the single-band time delays between macroimages of the PG 1115+080 and HE 0435-1223 gravitationally lensed quasars. In Fig. \ref{polynomial approximations} we show the polynomial approximations provided by our method for the light curves in Fig. \ref{light curves}: the curves are reduced to the same magnitude level, and the first-order terms are eliminated.

An additional factor restricting the accuracy of determining time delays is the character of the intrinsic quasar variability. In order to be suitable for determining the time delays, the intrinsic quasar variability curves must contain features with a characteristic timescale shorter than (or at least on the order of) the expected value of the time delay, and with the amplitudes exceeding the photometry errors inherent in observations. Unfortunately, this is not always the case with Q2237+0305.  Figure \ref{light curves} shows that  the light curves of the 2004 and 2005 seasons in filter $R$ have a rather long time coverage and are sampled densely enough, with an almost daily cadence that is sometimes interrupted by short weather gaps. Although these curves do not fully meet the criteria indicated above, we decided to use them to determine the interband time delays. The $V$ and $I$  light curves shown in Fig. \ref{light curves} are represented by a smaller number of data points than those in $R$, but not so few as to abandon the attempt of measuring the interband time delays.

\subsection {Results and analysis of time delays}

The time delays for three pairs of filters, $\tau_{RV}$, $\tau_{IR}$
, and $\tau_{IV}$
calculated from the data of the 2004 and 2005 seasons are presented in Table \ref{time lags observers} for each
of the A, B, and C macroimages separately (columns 2, 4, and 6), as well as averaged over
the components (columns 3, 5, and 7). We did not use the light curves of the faintest D component because the photometric accuracy is low. The positive values of the time delays mean that the light curves corresponding to the first letter in a subscript lag those corresponding to the second symbol.

To estimate the errors of the time delays, we used the initial light curves modified
in the following way. Up to 30\% of the data points were excluded from an initial
pair of light curves sequentially in a random manner, then the procedure of determining the time delay was applied for each such new realization of the light-curve pairs, and the corresponding new estimate of the time delay was obtained.
A number of such trials reached 20, and then the average time-delay values were calculated for a set of such estimates. The value of the
RMS deviation from the average was taken as an error of our time-delay estimates.
This approach to estimating the errors of measuring the time delays differs from
the generally accepted approach. A pair of artificial model signals is usually
synthesized, with the subsequent signal imposing various random noise realizations that imitate the scatter of the initial data points. Our  approach is also valid. At least, it has the advantage that we need not
worry whether the excepted random noise parameters (e.g., variance and
probability density distribution) are adequate for the real characteristics of the photometry
errors of the processes under comparison.

%***************************Figure  **********************************************
%**************************************************************************************

\begin{figure*} [!htp]
\resizebox{18.3 cm}{!}{\includegraphics{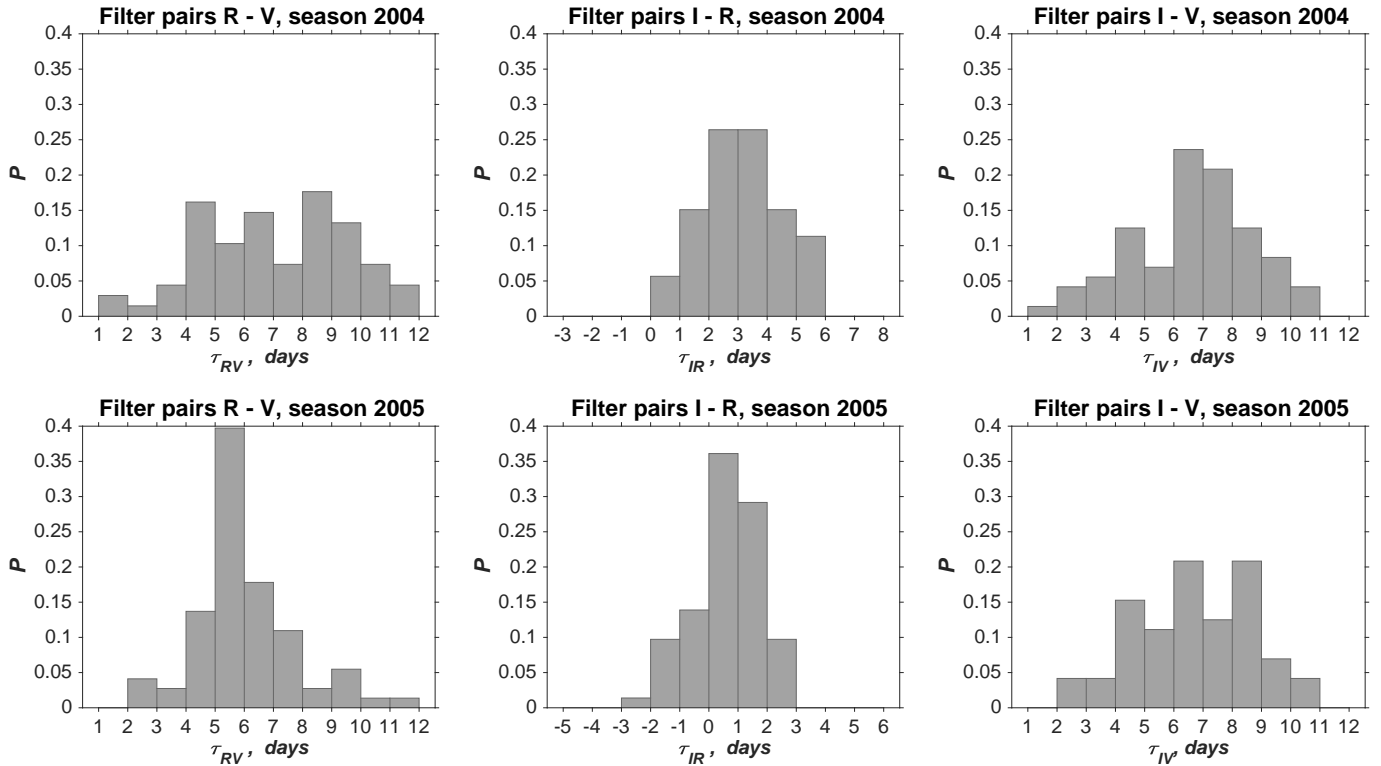}}
 \caption{ Histograms of the probability density distribution $P$ for the estimates of the interband time delays $\tau_{RV}$, $\tau_{IR}$ , and $\tau_{IV}$ presented in Table \ref{time lags observers}. }
 \label{histograms}
\end{figure*}

In Fig. \ref{histograms} we show the histograms, which demonstrate the
probability density distributions for the estimates of the time delays inherent in
our measurements presented in Table \ref{time lags observers}. The histograms, far enough from being Gaussian, are nevertheless symmetric enough and have fairly clear maxima, with the
exception of measurements for $\tau_{RV}$ in 2004 and, perhaps, in 2005.

Our measurements thus show quite definitely that the quasar intrinsic flux variations in
longer wavelengths lag those in shorter wavelengths for all the three pairs
of spectral bands, in accordance with the thermal reprocessing scenario proposed by
Krolik et al. (1991). For further consideration, we require that the time
delays are reduced to the source coordinate system, $\tilde \tau =\tau/(1+z_q)$. This
is shown in Table \ref{rest-frame time}, where the estimates of $\tilde \tau_{RV}$, $\tilde \tau_
{IR}$ , and $\tilde \tau_{IV}$ averaged over three macroimages are presented for two
seasons separately. The 80\%\ confidence intervals $(CI)$ for the estimated time
delays are also presented in this table.

%******************Table 2 **************************************************************************
\begin{table*} [!htp]
\caption {Relative interband time lags $\tau$ (in days) of the Q2237+0305 quasar for the $RV$, $IR,$ and $IV$ pairs of
light curves from the monitoring data by Dudinov et al. (2011) (in the observers's coordinate system).}
\label{time lags observers}
\centering
\setlength \extrarowheight{2pt}
\begin{tabular}{lcccccc}
\hline
 Season 2004  & $\tau_{RV}$ & $\tau^{ABC}_{RV}$&$\tau_{IR}$& $\tau^{ABC}_{IR}$&
$\tau_{IV}$&$\tau^{ABC}_{IV}$ \\ [3pt]
\hline
Comp A &$9.14\pm2.51$&             &$2.21\pm1.07$&             &$5.67\pm2.21$&            \\
Comp B &$6.68\pm2.31$&$7.18\pm2.81$&$3.89\pm1.08$&$3.18\pm1.37$&$6.81\pm1.91$&$6.75\pm2.34$\\
Comp C &$5.61\pm2.38$&             &$4.49\pm0.53$&             &$7.76\pm2.37$&            \\
\hline
Season 2005 & $\tau_{RV}$ & $\tau^{ABC}_{RV}$&$\tau_{IR}$& $\tau^{ABC}_{IR}$&
$\tau_{IV}$&$\tau^{ABC}_{IV}$ \\[3pt]
\hline
Comp A &$7.31\pm2.06$&             &$0.63\pm0.83$&             &$7.69\pm1.39$&            \\
Comp B &$5.64\pm2.35$&$6.25\pm1.97$&$0.89\pm1.81$&$0.63\pm1.36$&$5.83\pm2.48$&$6.72\pm2.28$\\
Comp C &$5.82\pm0.62$&             &$0.39\pm1.19$&             &$6.57\pm2.42$&            \\
\hline
\hline
 Averaged &    & $\tau^{ABC}_{RV}$   &   & $\tau^{ABC}_{IR}$ &   &   $\tau^{ABC}_{IV}$  \\[3pt]
\hline
Comp ABC &    & $6.71\pm2.46$ &    &   $1.71\pm1.86$   &   &   $6.73\pm2.31$\\
\hline
\end{tabular}
\end{table*}
%********************************************************************************************************

We consider the time delays for the 2004 and 2005 seasons represented in Table \ref{rest-frame time} separately. We further used their values averaged over the seasons, which are (in days)

\begin{equation}
\tilde \tau_{RV}=2.49 \pm 0.92; \, \, \,
\tilde \tau_{IR}=0.64 \pm 0.69; \, \, \,
\tilde\tau_{IV}=2.51 \pm 0.86.
\end{equation}

The longest (and almost equal) time delays were obtained for the $RV$ and $IV$
pairs of light curves, although the wavelength bases for the filter pairs $I-V$ and $R-V$ differ by four times. The least value of the interband time delay was
obtained for the intermediate wavelength base (and for the longest wavelength range of the filter pairs), namely, between the light curves in filters
$I$ and $R$.  This result is inconsistent with the expected power-law dependence of the disk effective radius on wavelength, $r_{\lambda_1}= r_{\lambda_0}({\lambda_1}/{\lambda_0})^\zeta$, with $\zeta=4/3$ for the standard thin accretion disk model. In the framework of the hypothesis that the radiation from the central region is reprocessed in its propagation toward the accretion disk periphery, there should be delays in the response at different wavelengths, which for the disk zones emitting at the effective wavelengths $\lambda_1$ and $\lambda_0$ can be written as

\begin{equation}
  \tau = \frac{r_{\lambda_0}}{c} \left[\left(\frac{\lambda_1}{\lambda_0}\right)^{\zeta}-1\right],
\end{equation}
where $c$ it is the velocity of light.

We used here a somewhat simplified scheme for emerging reverberation responses, which admits that they are formed for each of the filters in some annular
zones of an accretion disk. The zones are located at the distances from the central
source where the temperatures match the passbands of the corresponding filters. An
initial signal (fluctuations of the hard radiation from the central source region) in its
propagation toward the disk periphery is reprocessed into the longer-wavelength
signals with the time lags determined by the proper light travel times. At the same
time, the initial signal undergoes distortions, which are due to sizes, shapes, and positions of the
re-emitting regions. According to our simplified scheme, the observed distortion of
the initial signal in time (transfer function, or, more exactly, response function)
is determined in particular by the width and brightness profile of a specific
emitting zone, azimuthal distribution of the zone surface brightness, and finally,
by the inclination of the disk plane with respect to the plane of the sky.

%******************Table 3******************************************************************************
\begin{table}
\caption {Rest-frame time delays $\tilde \tau $ of the Q2237+0305 quasar for the $RV$,
$IR,$ and $IV$ pairs of light curves averaged over the three components, and the
corresponding confidence intervals $CI$ for the 80\% confidence level. }
\label{rest-frame time}
\centering
\setlength \extrarowheight{2pt}
\begin{tabular}{lccc}
\hline
Season 2004 & $\tilde \tau_{RV}$& $\tilde \tau_{IR}$&$\tilde \tau_{IV}$ \\
\hline
$\tilde \tau_{ABC}$ (days) & $2.67\pm1.04$  &$1.18\pm0.51$ & $2.44\pm0.87$ \\
$CI$      & (1.33; 4.0) & (0.53; 1.84) & (1.33; 3.56) \\
\hline
Season 2005 &$\tilde \tau_{RV}$& $\tilde \tau_{IR}$&$\tilde \tau_{IV}$ \\
\hline
$\tilde \tau_{ABC}$ (days) & $2.32\pm0.73$ &$0.23\pm0.5$ & $2.5\pm0.85$ \\
$CI$& (1.38; 3.26) & (-0.41; 0.88)  & (1.41; 3.58)\\
\hline
\hline
Averaged & $\tilde \tau^{ABC}_{RV}$   &   $ \tilde \tau^{ABC}_{IR}$ &      $ \tilde  \tau^{ABC}_{IV}$  \\
\hline
$\tilde \tau_{ABC}$ (days) & $2.49 \pm 0.92$ &      $0.64 \pm 0.69$   &      $ 2.51 \pm 0.86 $\\
\hline
\end{tabular}
\end{table}
%**************************************************************************************************

In each specific case, all these factors are unknown, or are known with a
high degree of uncertainty. To construct the response function, model
representations are therefore commonly used, which are simplified as a rule. In our case, the
simplification implies that the widths and brightness of the annular zones do not
change in azimuth. The width of the response function is then determined by a
temperature profile of the annular zone, which matches the corresponding filter
passband, while the effect of the disk inclination is reduced to additional broadening
of the response function by $\pm$ ($r$ sin $i$) / $c$, where $i$ is the inclination
angle, and $c$ is the velocity of light. It is important that the broadening is symmetric
with respect to the initial signal arrival time at a distance $r$ from the source.

Poindexter \& Kochanek (2010) have concluded from the analysis of
microlensing events in Q2237+0305 that the accretion disk in this quasar is observed
virtually face-on (cos $i$=0.66).
It is easy to calculate that for this inclination, broadening of the response function
for the zone radius, for example, $r=2\cdot 10^{16}$ cm, will be approximately $\pm 2$ days. The
effect of a response function with such a width on the initial signal with a
characteristic variability timescale of a few dozen days and longer will result only
in its insignificant smoothing, while the symmetric nature of the response function
will ensure the absence of biases in estimates of the interband time delays.

\section{Discussion}

    \subsection {Comparison with the classical thin-disk model}

For the standard thin accretion disk by Shakura \& Sunyaev (1973), the radius $r_\lambda$ (the disk scale length) where the disk temperature reaches
the photon energy, $kT =hc/\tilde\lambda$, is given by the expression (Poindexter \&
Kochanek 2010; Frank et al. 2002)

%***************************Figure ********************************************************
%**************************************************************************************

\begin{equation}
r_{\tilde\lambda}=\left(\frac{45G}{16\pi^6 hc^2}\right)^{1/3}\tilde{\lambda}^{4/3} (M_{BH}
\dot M)^{1/3},
\end{equation}
where $G$ is the gravitational constant, $h$  is the Planck constant, $c$ is the light speed,
$\tilde\lambda$ is the rest-frame wavelength, and $M_{BH}$ and $\dot M$ are the black hole mass and
accretion rate, respectively.

Poindexter \& Kochanek (2010) and  Morgan et al. (2010) proposed the following form of
expression (3):
\begin{equation}
r_\lambda=9.7 \cdot 10^{15}\left(\frac{\tilde\lambda}{1000}\right)^{4/3} \left(\frac
{M_{BH}}{10^9 M_\odot}\right)^{2/3}\left(\frac {L}{\eta L_{Edd}}\right)^{1/3},
\end{equation}
where dimensionless quantities $M_{BH}/10^9 M_\odot$ and $\tilde\lambda/1000\mu m$ are
introduced, $L$ and $L_{Edd}$ are the disk luminosity and the Eddington luminosity limit, $L_{Edd}=
10^{38}\cdot (M_{BH}/M_\odot)$ erg/s, and $\eta$ is the accretion efficiency, which
varies from 0.06 for the Schwarzshild black hole to 0.4 for the Kerr black hole. Frank et al. (2002) proposed considering $\eta=0.1$ as a reasonable value when
other information is absent.  $L$ is also often assumed to be equal to $L_{Edd}$, thus providing the upper limit to the estimate of $r_\lambda$.

For Q2237+0305, several $M_{BH}$
determinations are available in the literature. We have collected most of them  in Table \ref{black hole mass} and indicate the method used for the determination. The estimates vary almost
by an order of magnitude depending on the method. For example, adopting $M_{BH}=9\cdot10^8 M_\odot$  by Morgan et al. (2010), we obtain from Eq. (4) the following values of
radii (in centimeters) for the rest-frame wavelengths corresponding to filters $V$,
$R,$ and $I$:

\begin{equation}
\begin{split}
\quad \quad\quad
 & r_V=(2.32\pm0.32)\cdot10^{15},{} \\
& {} r_R=(2.84\pm0.40)\cdot 10^{15}, {} \\
& {} r_I=(4.37\pm0.60)\cdot10^{15}.
\end{split}
\end{equation}

These values are generally consistent with the determinations of the effective accretion
disk size in Q2237+0305 made from observations of microlensing events.
Most of them are collected in Table \ref{accretion disk}, where $r_{1/2}$ is the so-called half-light
radius equal to $2.44 r_\lambda$ (Poindexter \& Kochanek 2010).

Accordingly, we may expect the following quantities for the distances between the disk
zones that radiate in the corresponding spectral bands:

\begin{equation}
\begin{split}
\quad \quad\quad
 & r_{R}-r_V=(0.52\pm0.64)\cdot10^{15},{} \\
& {} r_I-r_R=(1.53\pm0.80)\cdot10^{15}, {} \\
& {}  r_I-r_V=(2.05\pm1.20)\cdot10^{15}.
\end{split}
\end{equation}

Using a simplified model for the occurrence of the reverberation responses described above, we therefore obtain the following values of the interband time delays (in
days) predicted by the standard thin disk model:

\begin{equation}
\begin{split}
\quad \quad\quad
 & \tilde\tau_{RV}=0.2\pm0.26,  {} \\
& {}  \tilde\tau_{IR}= 0.6\pm0.31, {} \\
& {} \tilde\tau_{IV}= 0.8\pm0.47.
\end{split}
\end{equation}

%******************Table 4*****************************************************
\begin{table*} [!htp]
\caption {Estimates of the black hole mass in the Q2237+0305 quasar collected from the works published
during 2000 - 2015.}
\label{black hole mass}
\centering
\setlength \extrarowheight{2pt}
\begin{tabular}{lcl}
\hline
Authors             &  $M_{BH}\cdot 10^8 M_\odot$&  Methods and details\\ [2pt]
\hline
Agol et al. (2000)&20 & From the flux ratios in the mid-IR\\
Kochanek (2004)     &  3.3  & From $M_{BH}$ - $L$ relationship assuming $\eta= 0.15$, $L = L_{Edd}$\\
Pooley et al. (2007)& 10    & From bolometric luminosity assuming $\eta = 0.15$, $L =L_{Edd}$\\
Morgan et al. (2010)&  9    & From C {\scriptsize IV} velocity width by Yee \& De Robertis (1991)\\
Assef et al. (2011) &  6.17 & From C {\scriptsize IV}, H {\scriptsize $\alpha$} and H {\scriptsize $\beta$} velocity widths\\
Sluse et al. (2011) &  2.0  & From C {\scriptsize IV} velocity width from Assef et al. (2011)\\
Sluse et al. (2011) & 17.7  & From the disk radius - $M_{BH}$ relationship\\
Mediavilla et al. (2015a, 2015b) & 12& From a central depression in microlensing light curve\\
\hline
\end{tabular}
\end{table*}
%********************************************************************************************

Uncertainties in quantities (5), (6), and (7) were calculated adopting the formal uncertainty in the $M_{BH}$  determination from the line widths to be equal to approximately 0.1 dex (Morgan et al. 2010). It should be noted that some authors have reported less optimistic errors for the broadline and reverberation mapping estimates of $M_{BH}$  (e.g., Kollmeier et al. 2006; Kelly et al. 2008).

The interband time delays calculated with the use of Eq. (4) are noticeably shorter
than those determined in the present work (Table \ref{rest-frame time} and the quantities in Eq. (1) averaged over the two seasons). The discrepancy is large enough and needs to be discussed.

\subsection{The problem of large accretion disk radii}

We are not the first to encounter this discrepancy between the measured reverberation time lags and those predicted by the standard accretion disk model. In particular, Morgan et al. (2010), Edelson et al. (2015), Kokubo (2018), and the DES and STORM project participants (Fausnaugh et al. 2016, 2018; Mudd et al. 2018) pointed out that the measured time delays between the quasar flux variations in different spectral regions are often noticeably longer than the expected values derived within the standard model of a thin accretion disk. The evident explanation of this discrepancy might imply that the observed reverberation responses arise in some extended disk regions located somewhere at the disk periphery.

The idea of the existence of such a structure has repeatedly been expressed in a number of earlier works dedicated, for instance, to the analysis of flux ratio anomalies, or interpreting the results of microlensing studies in some gravitationally lensed quasars (Witt \& Mao
1994; Pooley et al. 2007; Vakulik et al. 2007; Shulga et al. 2014). The similar
discrepancy between the measured dimensions of quasars and those derived from their
luminosities has also been discussed by Dai et al. (2003), Poindexter et al. (2008), Morgan et al. (2010), and Poindexter \& Kochanek (2010). Jaroszynski et al. (1992) were the first to admit the existence of an extended outer structure in
the Q2237+0305 accretion disk in their simulations of microlensing events.
 They noted that such a structure could add 100\% more light to the thin-disk radiation in $B$ and
$R$. They suggested that this additional light could be the radiation from the disk reprocessed in the outer regions of the quasar. This extended structure does not experience microlensing, but damps the microlensing peaks produced by the compact inner parts of the disk.  Schild \& Vakulik (2003) and Vakulik et al. (2007) have shown that a two-component quasar structure model consisting of a central compact source and an extended outer feature is capable of better reproducing in simulations the observed amplitudes of microlensing light curves than the central source alone.

The accretion disk models predict a power-law dependence of the disk effective radius on wavelength, $r_\lambda \sim \lambda^\zeta$ , with $\zeta = 4/3$ for the standard thin accretion disk. For Q2237+0305, there are several determinations of  the exponent values based on observations of chromatic microlensing events, which are rather well (Eigenbrod et al. 2008) or marginally (Anguita et al. 2008; Munoz et al. 2016) consistent with the thin-disk model predictions. However, an analysis of multiwavelength data for 12 objects presented by Blackburne et al. (2011) not only confirmed the abnormally large disks for most of the objects, but also demonstrated  a noticeable diversity of the disk radii dependences on wavelength.

\subsection{Scattering envelope around the black hole?}

These discrepancies between the results of observations and predictions made from a standard disk model are important indications that the real accretion disk may noticeably differ from the model. A possible scenario for the emergence of an extended structure around the accretion disk was
 considered several decades ago in the classical work by Shakura \& Sunyaev
(1973). They analyzed a supercritical accretion mode, which is
characterized by the luminosity fixed at the Eddington limit $L_{cr}=10^{38}M /
M_{\odot}$ erg/s and by a high accretion rate, $\dot M > \dot M_{cr}$, where $\dot
M_{cr}= L_{cr}/\eta c^2$, as well as by a low value of the efficiency $\alpha$ of
the angular momentum transport in the accreting matter, $\alpha \ll 1$. According
to the classical accretion disk model, the disk thickness increases with distance
to the black hole, providing the possibility that the outer regions of the disk
intercept a portion (from 0.1 to 10\%) of the hard radiation of the central regions
and re-emit it in the UV and optical continuum, as well as in emission lines of
various elements. Shakura \& Sunyaev (1973) showed that in the supercritical
regime, when $\dot M\gg \dot M_{cr}$ and $\alpha \ll 1$, an optically thick
scattering envelope is formed, which increases the apparent disk size.
They gave analytical expressions for the effective temperature $T_{eff}$ and radius
$r_{eff}$ of such an envelope,

\begin{equation}
T_{eff} \simeq 2\cdot10^{10}\dot m^{-15/11}m^{-2/11}\alpha^{10/11}A^{-6/11} (^\circ K)
\end{equation}

\begin{equation}
r_{eff}\simeq 3\cdot10^{2}\dot m^{51/22}m^{10/11}\alpha^{-17/11}A^{8/11} (cm)
.\end{equation}

%******************Table 5*******************************
\begin{table*} [!htp]
\centering
\caption {Half-light radius $r_{1/2}$ of the Q2237+0305 quasar
accretion disk determined in different works from the analysis of microlensing events.
To compare this with the scale length $r_\lambda$, the relation $r_{1/2}=2.44 \, r_\lambda$
should be used.}
\label{accretion disk}

\setlength \extrarowheight{2pt}
\begin{tabular}{p{5cm}p{6cm}c}
\hline
Authors & Spectral range & $r_{1/2}$ (cm)  \\ [3pt]
\hline
Kochanek (2004)          & Johnson-Cousins $V$ filter              &$3\cdot 10^{15}$    \\
Wayth et al. (2005)       & C {\scriptsize III]} and Mg {\scriptsize II} emission lines      &$\leq 6.2\cdot 10^{16}$ \\
Vakulik et al. (2007)     & Johnson-Cousins $V$ filter              & $2.34\cdot 10^{15}$\\
Anguita et al. (2008)     & SDSS $g$` and $r$` + Johnson-Cousins $V$       & $1.57\cdot 10^{15}$\\
Morgan et al. (2010)     & Johnson-Cousins $V$ filter              & $4.7\cdot 10^{15}$ \\
Poindexter \& Kochanek (2010)& Johnson-Cousins $V$ filter            & $1.41\cdot 10^{16}$\\
Mosquera et al. (2013)   & Johnson-Cousins $V$ filter              & $2.57\cdot 10^{16}$ \\
Mosquera et al. (2013)   & Soft X-rays                           & $5.75\cdot 10^{15}$ \\
Mosquera et al. (2013)   & Hard X-rays                           & $2.88\cdot 10^{15}$ \\
Mediavilla et al. (2015a) & Johnson-Cousins $V$ filter              & $7.7\cdot 10^{15}$  \\
Mediavilla et al. (2015b) & Johnson-Cousins $V$ filter              & $1.09\cdot 10^{16}$\\
Munoz et al. (2016)      & 7 narrow filters + Bessel $I$, 3510 - 8130 A  & $2.2\cdot 10^{16}$ \\
Vives-Arias et al. (2016) & Mid-IR (10.36 $\mu m$)                 & $8.8\cdot 10^{15}$  \\
\hline
\end{tabular}
\end{table*}
%*************************************************************************************************************

Here, dimensionless parameters are introduced: $m=M/M_\odot$,  $\dot m=\dot M/
\dot M_{cr}$. Parameter $A$ characterizes a ratio of energy losses in the Compton
processes to those in the free-free transitions. For physical conditions of interest
in this particular considerations, $A$ is noted to vary from 10 to 300 (Shakura \&
Sunyaev 1973). They also pointed out that the temperature $T_{eff}$ is virtually
constant in the regions with $r > r_{eff}$. The authors further point out that in the optical wavelengths which correspond to low-frequency spectral range at the discussed temperature $T_{eff}$ the radius near which the envelope becomes opaque exceeds $r_{eff}$ considerably. Setting the value of the optical opacity
equal to unity, the authors obtained the following expression for the optical envelope radius (in centimeters):
\begin{equation}
r_{opt}\simeq10^7 \alpha^{-3/4}\left (\frac {10^{6\, \circ}K}{T}\right )^{3/8}
\left (\frac {10^{15}Hz}{\nu}\right )^{1/2}\dot m^{9/8}m^{3/4},
\end{equation}
where $(10^{6\,\circ}K)/T$ and $(10^{15}Hz)/\nu$ are dimensionless temperature and
frequency, respectively.

In Table \ref{two sets parameters}  we show the values of $T_{eff}$,
$r_{eff}$ , and $r_{opt}$ that were calculated for two pairs of parameters $\alpha$ and
$A$ selected arbitrarily within the range of their permissible values indicated above,
namely, for $\alpha=0.05$ and $A=100$, and $\alpha=0.1$ and $A=50$. The black hole mass
was taken from Table \ref{black hole mass} to equal $M_{BH}=9\cdot10^8 M_\odot$, and the dimensionless
accretion rate was adopted to be $\dot m=17$, according to Morgan et al. (2010) and
Abolmasov \& Shakura (2012). Table \ref{two sets parameters} shows that the supercritical regime provides values of $r_{opt}$ that are steadily higher than those predicted by the standard thin-disk model. The reverberation signals at these radii can therefore be expected to exhibit longer time lags.

\begin{figure*} [!htp]
\resizebox{18.3 cm}{!}{\includegraphics{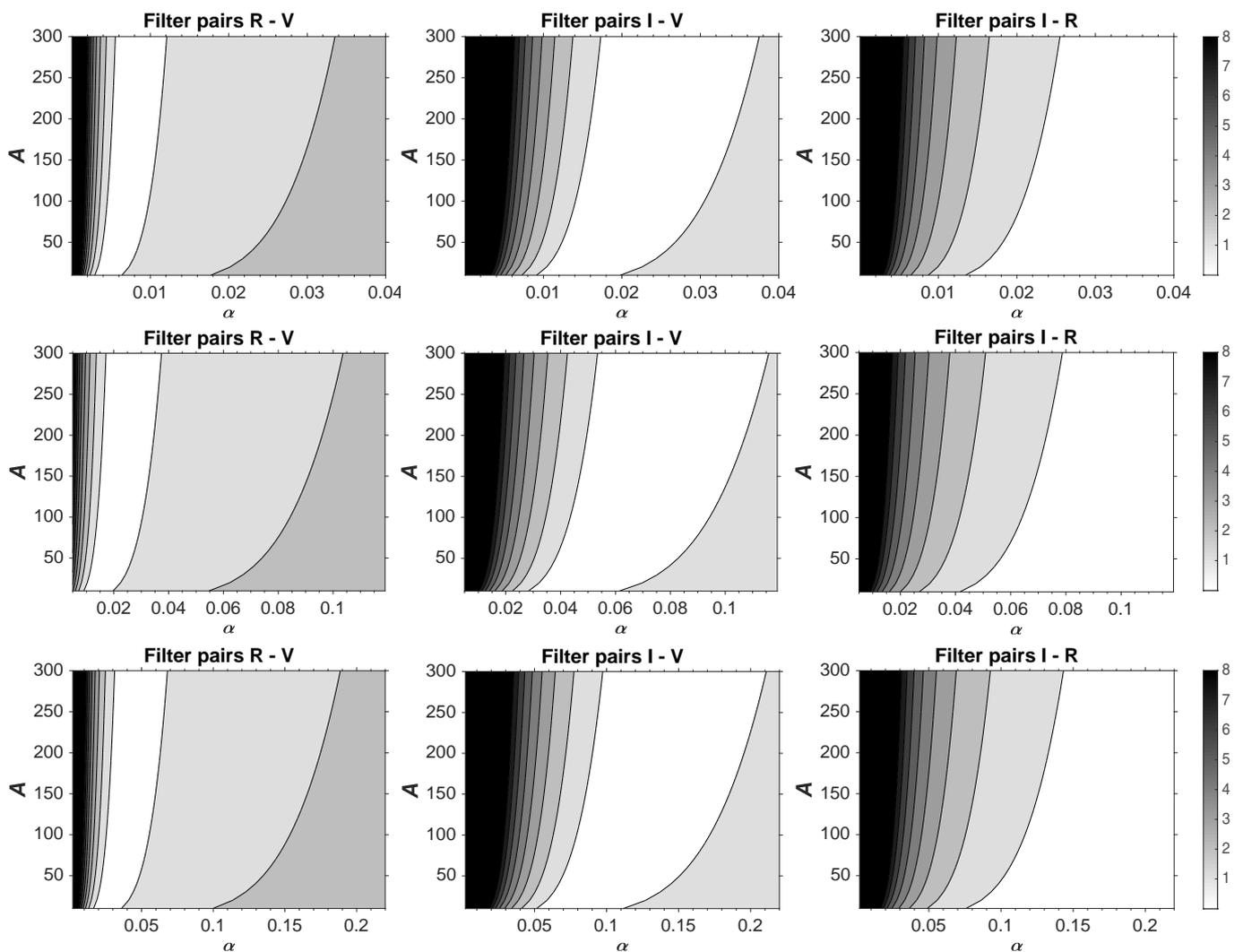}}
 \caption{Difference maps between the interband time delays determined in the present work and calculated according to equations (2) and (10). The brightest regions correspond to the smallest difference between these values. The maps are built for three black hole masses, from top to bottom: $M_{BH}=2\cdot10^8 M_\odot$, $M_{BH}=9\cdot10^8 M_\odot$ , and $M_{BH}=20\cdot10^8 M_\odot$. }
\label{maps all masses}
\end{figure*}

To reveal the regions of parameters $\alpha$ and $A$ that provide the best fit between the time delays determined in our work and those calculated for the radii $r_{opt}$ according to expression (10), we plot the difference maps between the measured and predicted values of the interband time lags.
The maps built in coordinates $\alpha$ and $A$ for each of the three filter pairs separately are shown in Fig. \ref{maps all masses}. The difference values are reproduced in grades of gray shown at the right edge of each row. The brightest regions correspond to the smallest difference between the reverberation time delays calculated according to Eq. (10) and those measured in the present work. Rather vast areas in coordinates $\alpha$ and $A$ provide the values of the interband time delays predicted with the use of Eq. 2 and Eq. 10 consistent with our measurements. These areas are clearly seen to shift toward the higher values of $\alpha$  for the longer wavelength range. This is consistent with the analysis of the supercritical regime by Shakura \& Sunyaev (1973), who noted that parameter $\alpha$ can (and must) depend on the disk radius. In particular, for the turbulent mechanism of the angular momentum transport, $\alpha$ can be about 1 at the disk periphery, while closer to the black hole, where the accretion picture becomes spherical, parameter $\alpha\sim 10^{-3}$ .

Another quantity that may serve as an indicator of the accretion regime in a particular object is the Eddington ratio, $L_{bol}/L_{Edd} $.  We followed Agol et al. (2009) and adopted $L_{bol} = 4 \cdot 10^{46}$   erg$\cdot$s$^{-1}$ for the bolometric luminosity of Q2237+0305 and calculated $ L_{Edd} =1.3 \cdot 10^{38}(M_{BH}/M_\odot$)    erg$\cdot$s$^{-1}$ for the black hole mass $M_{BH} = 9 \cdot 10^{8}M_\odot $
 (Morgan et al. 2010, see also Table \ref{black hole mass} ). We obtained $L_{bol}/L_{Edd} \approx 0.34$.

Collin et al. (2002) have analyzed the statistical relation between the Eddington ratios  $L_{bol}/L_{Edd} $  and black hole masses for 34 objects from the Kaspi et al. (2000) sample of the PG quasars for which the black hole masses are known. They reported that 11 objects have super-Eddington luminosities, and 7 have luminosities higher than $0.3\cdot L_{Edd}$, that is, more than a half of the objects are inconsistent with the geometrically thin-disk model.

Similar studies have been carried out by Kollmeier et al. (2006) using 407 AGNs from the AGN and Galaxy Evolution Survey (AGES). Their $L_{bol}/L_{Edd} $ -- $M_{BH}$ plot exhibits condensation of the data points (a ridge of a sort) at $L_{bol}/L_{Edd}\approx 1/4 $,  and demonstrates quite convincingly the lack of AGNs at a factor of approximately 10 below Eddington. They note that this can be regarded as strong evidence that supermassive BHs gain most of their mass while radiating close to the Eddington limit. The distributions of $L_{bol}/L_{Edd} $  built both at fixed BH mass and at fixed luminosity are log-normal and have rather sharp peaks near $L_{bol}/L_{Edd}\approx 1/4 $, thus confirming that the sample consists mostly of AGNs radiating near the Eddington limit. With these two works taken into account, we conclude that the Q2237+0305 quasar can be considered as an object radiating fairly near the Eddington limit.

%***************************Figure 3 ********************************************************
%**************************************************************************************

\section{Conclusions}

The scenario that an extended optically thick envelope forms in the supercritical accretion regime considered by Shakura \& Sunyaev (1973) therefore is in principle capable of explaining the high values of the interband time delays obtained in our work. The analysis of the supercritical regime and of its possible observational manifestations developed further in Abolmasov \& Shakura (2012) and Shakura (2018) has also been considered in simulations (e.g., Ohsuga et al. 2005; Okuda et al. 2005; Volonteri et al. 2015; Sakurai et al 2016; Jiang et al. 2019; Pinto et al. 2020, and others). The results of some observations have also made their authors consider the suggestion that some quasars may accrete in the super-Eddington regime (Collin et al. 2002; Kollmeier et al. (2006); Lauzuisi et al. 2016; Jin et al. 2017; Liu et al. 2019).

As we noted in Sec. 2, the behavior of our estimates of the interband time delays in wavelengths is not only inconsistent with the theoretic expectations, but is also very strange: almost equal $\tau_{RV}$ and $\tau_{IV}$ and very short $\tau_{IR}$. Although the uncertainties of determining the time lags presented in Table \ref{rest-frame time}  are large enough, we would like to draw attention to the low $\tau_{IR}$ value. Distracting from a possible physical reason, we note that if the values of $\tau_{RV}$ and $\tau_{IV}$ are really close to each other, the values of $\tau_{IR}$ must be low indeed, that is, the low $\tau_{IR}$  estimate can be regarded as an argument that $\tau_{RV}$ and $\tau_{IV}$ are measured correctly.

%******************Table 6*******************************
\begin{table} [!htp]
\centering
\caption{ $T_{eff}$, $r_{eff}$ , and $r_{opt}$ (in
centimeters) calculated from Eqs. (8), (9), and (10) for two sets of parameters
$\alpha$ and $A$. The black hole mass was adopted to be $M_{BH}=9\cdot 10^8
M_{\odot}$ (see Eq.  \ref{black hole mass}) and the dimensionless accretion rate was $\dot m=17$ according
to Morgan et al. (2010) and Abolmasov \& Shakura (2012).}
\label{two sets parameters}
\setlength \extrarowheight{2pt}
\begin{tabular}{llll}
\hline
Parameters $\alpha$, A & $T_{eff}$ $(^\circ K)$& $r_{eff}$ (cm) & $r_{opt}$ (cm) \\
\hline
$\alpha=0.1$; \,  $A=50$ &$1.4\cdot 10^{5}$&$3.6\cdot 10^{15}$&$1.3\cdot 10^{16}$\\
$\alpha=0.05$; $A=100$ &$0.5\cdot 10^{5}$&$2.3\cdot 10^{14}$&$3.2\cdot 10^{16}$\\
\hline
\end{tabular}
\end{table}
%**************************************************************************************

The time delays $\tau_{RV}$, $\tau_{IR}$ , and $\tau_{IV}$ measured in our work therefore appear to indicate that the disk radius ceased to depend on the wavelength after the zone in which the blackbody temperature corresponds to the passband of filter $R$. Theoretic analyses made by Shakura \& Sunyaev (1973), as well as presented later in Abolmasov \& Shakura (2012) and Shakura (2018) showed that scattering envelopes not only provide larger apparent disk sizes, but make their size less dependent on wavelength when scattering by free electrons is the main source of the envelope opacity. Numerical simulations also predict a constant temperature for the outer layers of the envelope, namely, for an optical depth less than 5 (Hubeny \& Hubeny 1998). Abolmasov \& Shakura (2012) also noted that an extended scattering envelope can be expected to spatially mix the radiation from different disk zones. These peculiarities of the envelope may qualitatively explain the abnormal behavior of our estimates of the interband time lags in wavelengths.

Before we summarise, we add one more comment about the high values of the reverberation time lags. As we indicated above, our measurements are inconsistent not only with the thin-disk model predictions, but also with the results of many microlensing observations of Q2237+0305. This is easily explained by the fact that microlensing events and reverberation responses concern different spatial scales of the accretion disks, with the smaller of them referring to microlensing. As numerical calculations by Jaroszynski et al. (1992) have shown, the caustic crossing events are most sensitive to the structure of the inner regions of the disk, while the extended outer regions may only contribute to the observed amplitudes of microlensing events.
We summarise our results below.

\begin{itemize}

\item [] We have measured the interband time delays between the light curves of the Q2237+0305 macroimages in filters $V$, $R,$ and $I$ to be $\tilde\tau_{RV}=2.49\pm0.92$; $\tilde\tau_{IR}=0.64\pm0.69,$ and $\tilde\tau_{IV}=2.51\pm0.86$ days (in the source coordinate frame), with flux variations in longer wavelengths lagging those in the harder range of the  spectrum, in accordance with the reprocessing scenario.

\item [] These delays indicate that the observed reverberation responses arise in structures located at distances from the central black hole that exceed the radii of the corresponding regions of the accretion disk predicted by the standard thin-disk model.

\item [] An explanation for this discrepancy is sought within the assumption about a super-Eddington accretion regime in quasar Q2237 + 0305, which leads to the formation of an extended scattering envelope (atmosphere) around the central black hole.

\item [] Based on a theoretical analysis of the super-Eddinton accretion and physical properties of the envelope reported by Shakura \& Sunyaev (1973), Abolmasov \& Shakura (2012), and Shakura (2018), we tested whether the observed reverberation responses might be formed in the corresponding zones of the envelope.

\item [] We showed that for certain values of the parameters $\alpha$ and $A$ (angular momentum transport efficiency and the ratio of energy losses in the Compton processes to those in the free-free transitions), the super-Eddington regime provides values of the envelope radius that are noticeably higher than those predicted by the standard thin-disk model. The reverberation signals at these radii can therefore be expected to exhibit longer time lags, consistent with our measurements.

\item [] To reveal the regions of parameters $\alpha$ and $A$ that provide the best fit between the interband time delays determined in our work and those based on the analysis of the super-Eddington accretion regime by Shakura \& Sunyaev (1973), we plotted the difference maps between these values in coordinates $\alpha$ and $A$ (Fig. \ref{maps all masses}). The areas of the best fit of our results to those expected from the Shakura \& Sunyaev (1973) analysis of the super-Eddington regime are clearly seen to shift towards the higher values of $\alpha$ for the longer wavelength range. This is consistent with their indication that parameter $\alpha$ can (and must) depend on the disk radius.

\item [] The Eddington ratio, $L_{bol}/L_{Edd}$, is also an important quantity that may serve as an indicator of the accretion regime in a particular object. When we adopt, according to Agol et al. (2009), $L_{Edd} =4 \cdot 10^{46}$ erg$\cdot$s$^{-1}$  for the Q2237+0305 bolometric luminosity and regard the black hole mass $M_{BH}=9\cdot10^8 M_\odot$  by Morgan et al. (2010) as known with an uncertainty of 0.3 dex, we obtain $0.26 < L_{bol}/L_{Edd} < 0.41$ for the possible value of the Q2237+0305 Eddington ratio. Comparing this with the results by Collin et al. (2002) and Kollmeier et al. (2006), who reported that most objects from their samples have Eddington ratios higher than 0.3 (Collin et al. 2002) or 0.25 (Kollmeier et al. 2006), we conclude that the Q2237+0305 quasar can be considered as an object that radiates fairly near the Eddington limit. It is therefore capable of developing an optically thick scattering envelope that changes the observed structural properties of the accretion disk.

\end{itemize}

\begin{acknowledgements}
The authors would like to thank the colleagues from IRA of the NAS of Ukraine and IA of the KhNU for the fruitful discussions of the problems raised in the present work. L. Berdina is grateful to the EAS/Springer (2018) and EPS (2019) grants which have provided an opportunity to report preliminary results of the present work at conferences. V. Tsvetkova would like to acknowledge the role of R.E. Schild  in giving rise to the RM studies in Kharkov, who tried to draw attention to the problem more than a decade ago. V. Shulga acknowledges the College of Physics, International Center of Future Science, Jilin University, China, for hosting during the time period of preparing the paper for publication. Also, the authors are thankful to the anonymous referee for the valuable comments and suggestions.
 \end{acknowledgements}


\begin{thebibliography}{71}

\bibitem [\protect\citeauthoryear{Abol}{2012}]{b1}
Abolmasov, P., \& Shakura, N.I. 2012, MNRAS, 427, 1867

\bibitem [\protect\citeauthoryear{Agol}{2000}]{b2}
Agol, E., Jones, B., \& Blaes, O. 2000, ApJ, 545, 657

\bibitem [\protect\citeauthoryear{Agol}{2009}]{b3}
Agol, E., Gogarten, S.M., Gorjian, V., \& Kimball, A. 2009, ApJ, 697, 1010

\bibitem [\protect\citeauthoryear{Angu}{2008}]{b4}
Anguita, T., Schmidt, R.W., Turner, E.L., et al. 2008, A\&A, 480, 327

\bibitem [\protect\citeauthoryear{Asse}{2011}]{b5}
Assef, R.J., Denney, K.D., Kochanek, C.S., et al. 2011, ApJ, 742, 93

\bibitem [\protect\citeauthoryear{Bach}{2009}]{b6}
Bachev, R.S. 2009, A\&A, 493, 907

\bibitem [\protect\citeauthoryear{Berd}{2018}]{b7}
Berdina, L.A., Tsvetkova, V.S, \& Shulga V.M. 2018, RPRA, 23, 235 [in Russian]

\bibitem [\protect\citeauthoryear{Berd}{2019}]{b8}
Berdina, L.A., Tsvetkova, V.S, \& Shulga V.M. 2019, RPRA, 24, 242 [in Russian]

\bibitem [\protect\citeauthoryear{Blackb}{2011}]{b9}
Blackburne, J., Pooley, D., Rappaport, S., \& Schechter, P.L. 2011, ApJ, 729, 34

\bibitem [\protect\citeauthoryear{Bland}{1982}]{b10}
Blandford, R. D., \& McKee, C.F. 1982, ApJ, 255, 419

\bibitem [\protect\citeauthoryear{Cack}{2007}]{b11}
Cackett, E.M., Horne, K., \& Winkler, H. 2007, MNRAS, 380, 669

\bibitem [\protect\citeauthoryear{Coll}{1999}]{b12}
Collier, S., Horne, K., Wanders, I., \& Peterson, B.M. 1999, MNRAS, 302, L24

\bibitem [\protect\citeauthoryear{Coll}{2001}]{b13}
Collier, S. 2001, MNRAS, 325, 1527

\bibitem [\protect\citeauthoryear{Coll}{2002}]{b14}
Collin, S., Boisson, C., \& Mouchet, M., et al. 2002, A\&A, 388, 771

\bibitem [\protect\citeauthoryear{Cris}{1997}]{b15}
Cristiani, S., Trentini, S., La Franca, F., \& Andreani, P. 1997, A\&A, 321, 123

\bibitem [\protect\citeauthoryear{Dai}{2003}]{b16}
Dai, X., Chartas, G., Agol, E., Bautz, M.W., \& Garmire, G.P. 2003, ApJ, 589, 100

\bibitem [\protect\citeauthoryear{Dudi}{2011}]{b17}
Dudinov, V.N., Smirnov, G.V., Vakulik, V.G., Sergeev, A.V., \& Kochetov, A.E. 2011, RPRA, 2, 115

\bibitem [\protect\citeauthoryear{Edel}{2015}]{b18}
Edelson, R., Gelbord, J.M., Horne, K., et al. 2015, ApJ, 806, 129

\bibitem [\protect\citeauthoryear{Eigen}{2008}]{b19}
Eigenbrod, A., Courbin, F., Meylan, G., et al. 2008, A\&A 490, 933

\bibitem [\protect\citeauthoryear{Edri}{2012}]{b20}
Edri, H., Rafter, S.E., Chelouche, D., Kaspi, S., \& Behar, E. 2012, ApJ, 756, 73

\bibitem [\protect\citeauthoryear{Faus}{2016}]{b21}
Fausnaugh, M.M., Denney, K.D., Barth, A.J., et al. 2016, ApJ, 821, 56

\bibitem [\protect\citeauthoryear{Faus}{2018}]{b22}
Fausnaugh, M.M., Starkey, D.A., Horne, K., et al. 2018, ApJ, 854, 107

\bibitem [\protect\citeauthoryear{Frank}{2002}]{b23}
Frank J., King A., \& Raine D. 2002, Accretion Power in Astrophysics, ed. Cambridge University Press, 384

\bibitem [\protect\citeauthoryear{Give}{1999}]{b24}
Giveon, U., Maoz, D., Kaspi, S., Netzer, H., \& Smith, P.S. 1999, MNRAS, 306, 637

\bibitem [\protect\citeauthoryear{Gopa}{2013}]{b25}
Gopal-Krishna, Joshi, R., \& Chand, H. 2013, MNRAS, 430, 1302

\bibitem [\protect\citeauthoryear{Grier}{2017}]{b26}
Grier, C.J., Trump, J.R., Shen, Y., et al. 2017, ApJ, 851, 21

\bibitem [\protect\citeauthoryear{Grier}{2019}]{b27}
Grier, C.J., Shen, Y., Horne, K. et al. 2019, ApJ, 887, 38

\bibitem [\protect\citeauthoryear{Homay}{2019}]{b28}
Homayouni, Y., Trump, J.R., Grier, C.J. et al. 2019, ApJ, 880, 126

\bibitem [\protect\citeauthoryear{Huben}{1998}]{b29}
Hubeny, I., \& Hubeny, V. 1998, ApJ, 505, 558

\bibitem [\protect\citeauthoryear{Huch}{1985}]{b30}
Huchra, J., Gorenstein, M., Kent, S., et al. 1985, AJ, 90, 691

\bibitem [\protect\citeauthoryear{Jaros}{1992}]{b31}
Jaroszynski M., Wambsganss J., \& Paczynski B. 1992, ApJ, 396, L65

\bibitem [\protect\citeauthoryear{Jiang}{2016}]{b31}
Jiang, L., Shen, Y., McGreer, I.D., et al. 2016, ApJ, 818, 137

\bibitem [\protect\citeauthoryear{Jiang}{2019}]{b33}
Jiang, Y.-F., Stone, J.M., \& Davis, S.W. 2019, ApJ, 880, 67

\bibitem [\protect\citeauthoryear{Jin}{2017}]{b34}
Jin, C., Done, C., \& Ward, M. 2017, MNRAS, 468, 3663\

\bibitem [\protect\citeauthoryear{Kaspi}{2000}]{b35}
Kaspi, Sh., Smith, P.S., Netzer, H., et al. 2000, ApJ, 533, 631

\bibitem [\protect\citeauthoryear{Kelly}{2008}]{b36}
Kelly, B.C., Bechtold, J., Trump, J.R., Vestergaard, M., \&  Siemiginowska, A. 2008, ApJS, 176, 355

\bibitem [\protect\citeauthoryear{Kinemu}{2020}]{b37}
Kinemuchi, K., Hall, P.B., McGreer, I., et al. 2020, ApJ, 250, 10

\bibitem [\protect\citeauthoryear{Koch}{2004}]{b38}
Kochanek, C.S. 2004, ApJ, 605, 58

\bibitem [\protect\citeauthoryear{Koku}{2018}]{b39}
Kokubo, M.  2018, Publ. Astron. Soc. Japan, 70, 97

\bibitem [\protect\citeauthoryear{Koku}{2020}]{b40}
Kokubo, M. \& Minezaki, T. 2020, MNRAS, 491, 4615

\bibitem [\protect\citeauthoryear{Koll}{2006}]{b41}
Kollmeier, J.A., Onken, C.A., Kochanek, C.S., et al. 2006, ApJ, 648, 128

\bibitem [\protect\citeauthoryear{Kopt}{2005}]{b31}
Koptelova, E., Shimanovskaya, E., Artamonov, B., et al. 2005, MNRAS, 356, 323

\bibitem [\protect\citeauthoryear{Kopt}{2006}]{b42}
Koptelova, E.A., Oknyanskij, V.L., \& Shimanovskaya E.V. 2006, A\&A, 452, 37

\bibitem [\protect\citeauthoryear{Kopt}{2010}]{b43}
Koptelova, E.A., Oknyanskij, V.L., Artamonov, B., \& Chen, W.-P. 2010, Mem. Soc. Astron. Ital., 81, 138

\bibitem [\protect\citeauthoryear{Krol}{1991}]{b44}
Krolik, J.H., Horne, K., Kallman, T.R., et al. 1991, ApJ, 371, 541

\bibitem [\protect\citeauthoryear{Kumar}{2015}]{b45}
Kumar, P., Gopal-Krishna, \& Chand, H. 2015,  MNRAS, 448, 1463

\bibitem [\protect\citeauthoryear{Lanz}{2016}]{b46}
Lanzuisi, G., Perna, M., Comastri, A., et al. 2016, A\&A, 590, A77

\bibitem [\protect\citeauthoryear{Liu}{2019}]{b47}
Liu, H., Luo, B., Brandt, W.N., et al. 2019, ApJ, 878, 79

\bibitem [\protect\citeauthoryear{Luo}{2020}]{b48}
Luo, Y., Shen, Y., \& Yang, Q. 2020, MNRAS, 494, 3686

\bibitem [\protect\citeauthoryear{Magd}{2006}]{b49}
Magdis, G., \& Papadakis, I.E. 2006, AGN Variability from X-rays to Radio Waves ASP Conference Series, 360, 37

\bibitem [\protect\citeauthoryear{Mediav}{2015a}]{b50}
Mediavilla, E., Jimenez-vicente, J., Munoz, J. A., \& Mediavilla, T. 2015a, ApJ, 814, 26

\bibitem [\protect\citeauthoryear{Mediav}{2015b}]{b51}
Mediavilla, E., Jimenez-Vicente, J., Munoz, J. A., Mediavilla, T., \& Ariza, O. 2015b, ApJ, 798, 138

\bibitem [\protect\citeauthoryear{Morg}{2010}]{b52}
Morgan, C.W., Kochanek, C.S., Morgan, N.D., \& Falco, E.E. 2010, ApJ, 712,1129

\bibitem [\protect\citeauthoryear{Mosq}{2013}]{b53}
Mosquera, A.M., Kochanek, C.S., Chen, B., et al. 2013, ApJ, 769, 53

\bibitem [\protect\citeauthoryear{Mudd}{2018}]{b54}
Mudd, D., Martini, P., Zu, Y., et al. 2018, ApJ, 862, 123

\bibitem [\protect\citeauthoryear{Munoz}{2016}]{b55}
Munoz, J.A., Vives-Arias, H., Mosquera, A.M., et al. 2016, ApJ, 817, 155

\bibitem [\protect\citeauthoryear{Ohsu}{2005}]{b56}
Ohsuga, K., Mori, M., Nakamoto, T., \& Mineshige, S. 2005, ApJ, 628, 368

\bibitem [\protect\citeauthoryear{Okuda}{2005}]{b57}
Okuda, T., Teresi, V., Toscano, E., \& Molteni, D. 2005, MNRAS, 357, 295

\bibitem [\protect\citeauthoryear{Pinto}{2020}]{b58}
Pinto, C., Mehdipour, M., Walton, D. J., et al. 2020, MNRAS, 491, 5702

\bibitem [\protect\citeauthoryear{Poind}{2008}]{b59}
Poindexter, S., Morgan, N., \& Kochanek, C.S. 2008, ApJ, 673, 34

\bibitem [\protect\citeauthoryear{Poind}{2010}]{b60}
Poindexter, S., \& Kochanek, C.S. 2010, ApJ, 712, 668

\bibitem [\protect\citeauthoryear{Pool}{2007}]{b61}
Pooley, D., Blackburne, J.A., Rappaport, S. \& Schechter, P.L. 2007, ApJ, 661, 19

\bibitem [\protect\citeauthoryear{Sakur}{2016}]{b62}
Sakurai, Y., Inayoshi, K., \& Haiman, Z. 2016, MNRAS, 461, 4496

\bibitem [\protect\citeauthoryear{Schmi}{2012}]{b63}
Schmidt, K.B., Rix, H.-W., Shields, J.C., et al. 2012, ApJ, 744, 147

\bibitem [\protect\citeauthoryear{Serg}{2005}]{b64}
Sergeev, S.G., Doroshenko, V.T., Golubinskiy, Y.V., Merkulova, N.I., \& Sergeeva, E.A. 2005, ApJ, 622, 129

\bibitem [\protect\citeauthoryear{Shak}{1973}]{b65}
Shakura, N.I., \& Sunyaev, R.A. 1973, A\&A, 24, 337

\bibitem [\protect\citeauthoryear{Shak}{2018}]{66}
Shakura, N., Accretion Flows in Astrophysics, 2018, ed. Springer, 419

\bibitem [\protect\citeauthoryear{Schil}{2003}]{b67}
Schild R., \& Vakulik V. 2003, AJ, 126, 689

\bibitem [\protect\citeauthoryear{Shul}{2014}]{b68}
Shulga, V.M., Minakov A.A., Vakulik V.G., Smirnov G.V., \& Tsvetkova V.S. 2014, Vol. 2. Dark matter: Astrophysical aspects of the problem, ed. Shulga V., 357

\bibitem [\protect\citeauthoryear{Slus}{2011}]{b69}
Sluse, D., Schmidt, R., Courbin, F.,; Hutsemékers, D., et al. 2011, A\&A, 528, 100

\bibitem [\protect\citeauthoryear{Tewe}{2013}]{b70}
Tewes, M., Courbin, F., \& Meylan, G. 2013, A\&A, 553, A120

\bibitem [\protect\citeauthoryear{Tsve}{2016}]{b71}
Tsvetkova, V.S., Shulga, V.M., \& Berdina, L.A. 2016, MNRAS, 461, 3714

\bibitem [\protect\citeauthoryear{Udal}{2006}]{b72}
Udalski, A., Szymanski, M.K., Kubiak, M., et al. 2006, Acta Astronomica, 56, 293

\bibitem [\protect\citeauthoryear{Vaku}{2004}]{b73}
Vakulik, V.G., Schild, R.E., Dudinov, V.N., et al. 2004, A\&A, 420, 447

\bibitem [\protect\citeauthoryear{Vaku}{2007}]{b74}
Vakulik, V.G., Schild, R.E., Smirnov, G.V., Dudinov, V.N., \& Tsvetkova, V.S. 2007, MNRAS, 382, 819

\bibitem [\protect\citeauthoryear{Vand}{2004}]{b75}
Vanden Berk, D.E., Wilhite, B.C., Kron, R.G., et al. 2004, ApJ, 601, 692

\bibitem [\protect\citeauthoryear{Vives}{2016}]{b76}
Vives-Arias, H., Munoz, J.A., Kochanek, C.S., Mediavilla, E., \& Jiménez-Vicente, J. 2016, ApJ, 831, 43

\bibitem [\protect\citeauthoryear{Vol}{2015}]{b77}
Volonteri, M., Silk, J., \& Dubus, G. 2015, ApJ, 804, 148

\bibitem [\protect\citeauthoryear{Wand}{1997}]{b78}
Wanders, I., Peterson, B. M., Alloin, D., et al. 1997, ApJS, 113, 69

\bibitem [\protect\citeauthoryear{Wayth}{2005}]{b79}
Wayth, R.B., O'Dowd, M., \& Webster, R.L. 2005, MNRAS, 359, 561

\bibitem [\protect\citeauthoryear{Webb}{2000}]{b80}
Webb, W., \& Malkan, M. 2000, ApJ, 540, 652

\bibitem [\protect\citeauthoryear{Wilh}{2005}]{b81}
Wilhite, B.C., Vanden Berk, D.E., Kron, R.G., et al. 2005, ApJ, 633, 638

\bibitem [\protect\citeauthoryear{Homa}{1994}]{b82}
Witt H.J., \& Mao S. 1994, ApJ, 429, 66

\bibitem [\protect\citeauthoryear{Yi}{2019}]{b83}
Yi, W., Brandt, W.N., Hall, P.B., et al. 2019, ApJ, 242, 28

\bibitem [\protect\citeauthoryear{Yu}{2020}]{b84}
Yu, Z., Martini, P., Davis, T.M., et al. 2020, ApJ, 246, 16


\end{thebibliography}
\end{document}